\documentclass{article} % \documentclass{} is the first command in any LaTeX code.  It is used to define what kind of document you are creating such as an article or a book, and begins the document preamble

\usepackage{amsmath} % \usepackage is a command that allows you to add functionality to your LaTeX code
\usepackage{authblk}
\usepackage{graphicx}

\title{Optimal Checkpoint Interval with Availability as an Objective Function} % Sets article title
\author{Nirmal Raj Saxena \and Saurabh Hukerikar \and Mikolaj Blaz \and Swapna Raj}
\affil{NVIDIA, Santa Clara, CA}
\date{\today} % Sets date for date compiled

\begin{document} % All begin commands must be paired with an end command somewhere
    \maketitle % creates title using information in preamble (title, author, date)

\begin{abstract}
    
    We present a simplified derivation of the optimal checkpoint interval in Young74 \cite{young_1974}.  The optimal checkpoint interval derivation in \cite{young_1974} is based on minimizing the total lost time as an objective-function. Lost time is a function of checkpoint interval ($T_c$), checkpoint save time ($T_s$), and average failure time ($T_f$). This simplified derivation yields lost-time-optimal $T_c = \sqrt{2T_f T_s}$, which is identical to the one derived in \cite{young_1974}.  For large scale-out super-computer or datacenter systems, what is important is the selection of optimal checkpoint interval that maximizes availability. Availability is a function of $T_f$, $T_s$, $T_c$ and checkpoint recovery time ($T_r$). We show that availability-optimal $T_c = T_s + \sqrt{2(T_f+T_r)T_s+{T_s}^2}$. For $T_s << T_r \And T_r << T_f$, availability-optimal checkpoint interval is asymptotically same as lost-time-optimal checkpoint interval. We show that these optimal checkpoint intervals hold in situations where the error detection latency $T_e$ is significantly smaller than any selected checkpoint interval. However, in cases where the error detection latency $T_e$ is very large then the optimal checkpoint interval is greater than or equal to $T_e$.
    
\end{abstract}
    
\section{Introduction} % creates a section

 For large scale systems comprising hundreds of thousands of compute nodes it is not unusual for $T_f$ to be in single digit hours. These failures are either due to transient faults (emission particle induced soft errors in memory elements or bit errors in communication links) or permanent faults in compute/networking nodes.  If the mission time of a single job running across all of these compute nodes is significantly smaller than $T_f$ then there is a high probability that job would complete without any error. However, some of the super-computing and datacenter applications have mission times larger than $T_f$. For example, the training time for GPT-3 \cite{Brown:2020} \emph{large language model} (LLM) in a 100 Petaflops/sec compute system would take 35 days.

With $T_f=1.0$ hour, there will be on average $840(=35\times24)$ fault interruption events during GPT-3 LLM training. Intuitively, if we pick a checkpoint interval every half hour ($T_c = \frac{T_f} {2}$) and ignoring the checkpoint save time ($T_s$), then on average about fifty percent of compute time will be lost or wasted. Young74 [1] derived an optimal checkpoint interval time $T_c = \sqrt{2T_f T_s}$, where $T_s$ is the exposed checkpoint state save time.  The derivation was based on minimizing the lost time using exponential failure rate density functions. This optimal checkpoint interval has been used by researchers and practitioners for the past fifty years and more recently in \cite{Du:2020} \cite{Daly:2006} \cite{Geist:2012} \cite{Cappello:2014} \cite{Snir:2014}. By exposed time, we mean the portion of context state save time that adds to the mission time overhead (due to synchronization primitives). For large scale systems, the entire checkpoint sate save is a background process which takes several minutes and minimizes the exposed time ($T_s$) which is typically in seconds.

There are three key contributions in this short note. Sections that follow cover these.

\begin{enumerate}

	\item 
	Using first order statistics formulation of the lost time as a function of $T_f, T_s, T_c \And T_r$, we re-derive the optimal checkpoint interval time $T_c = \sqrt{2 T_f T_s}$. This is identical to that in \cite{young_1974}.
	
	\item
	The most important performance metric in the data center is system availability. While minimizing lost time is an important contributor in maximizing availability, from the standpoint of optimal checkpoint interval the objective function that is relevant is availability. We show that availability-optimal $T_c = T_s + \sqrt{2(T_f+T_r)T_s+{T_s}^2}$. For $T_s << T_r \And T_r << T_f$, availability-optimal checkpoint interval is asymptotically same as lost-time-optimal checkpoint interval.
	
	\item 
	Factored in the derivation of optimal checkpoint interval is the implicit assumption that error detection latency (from the occurrence of a fault and its error response detection) is negligible compared to the checkpoint interval time. While this assumption is mostly true for some built-in error detection mechanisms (e.g. parity or ECC checkers that are close to memory structures), however, for other detection mechanisms like periodic diagnostics that detect errors long after the occurrence of a fault often results in $T_e >> T_c$. In these cases, we show that the optimal checkpoint interval is larger than the one without $T_e$ consideration.

\end{enumerate}

\section{Optimal Checkpoint Interval $T_c$}

\begin{figure} 
     \centering
     \includegraphics[width=\textwidth]{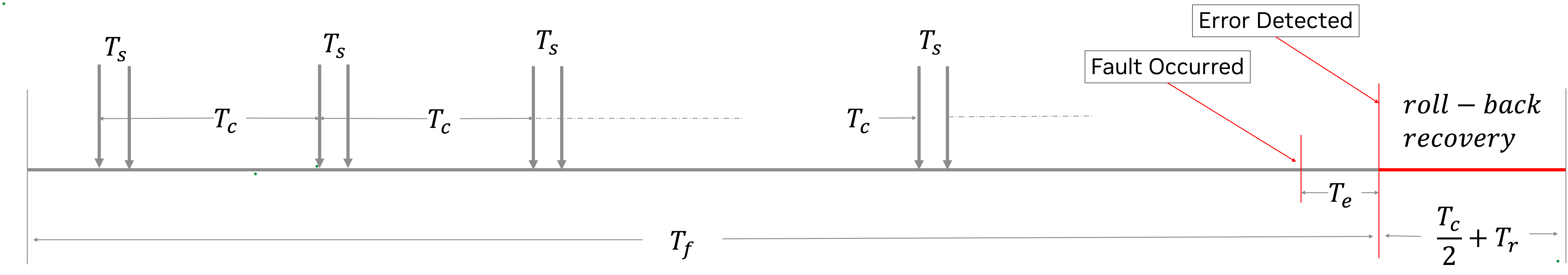}
     \caption{Timeline of First Fault Detection and Rollback Checkpoint Recovery}
     \label{fig:timeline}
\end{figure}

Figure \ref{fig:timeline} gives a timeline of events every $T_c$ interval with \emph{exposed} checkpoint state save time of $T_s$.  This timeline represents a mean time to failure ($T_f$) segment followed by, upon detection of error, checkpoint rollback recovery. The term 'exposed' is used here because typically checkpoint save is a background process. The only time that impacts the run-time of the mission application is the exposed time of the checkpoint state save that causes global synchronization. 

The total wasted or \emph{lost time} (in the sense no useful work is done by the mission application) for every recurrent failure event is given by

\begin{equation}
 LostTime(T_f, T_s, T_c, T_r) = \frac {T_f}{T_c} T_s + \frac{T_c} {2}+ T_r \label{eq:LT} \\
\end{equation}
The first term ($\frac{T_f T_s}{T_c}$) in Eqn. (\ref{eq:LT}) is the time it takes to save ($T_s$ time units) for $T_f/T_c$ checkpoint events. The other two terms are: the average checkpoint interval residual time ($T_c/2$) and the checkpoint state recovery time ($T_r$). Differentiating Eqn. (\ref{eq:LT}) with respect to $T_c$ we get Eqn. (\ref{eq:pLT}).

\begin{equation}
\frac {\partial LostTime(T_f, T_s, T_c, T_r)} {\partial T_c} = -\frac{T_f T_s} {{T_c}^2} + 1/2 = 0  \label{eq:pLT} \\
\end{equation}
Solving Eqn. (\ref{eq:pLT}) we get the optimal checkpoint interval in Eqn. (\ref{eq:optLT}), identical to the one derived in \cite{young_1974}.
\begin{equation}
T_c = \sqrt{2 T_f T_s} \label{eq:optLT} \\
\end{equation}

Availability is the fraction of time useful work is done by the mission application in a fault free interval. It is conventionally, the ratio of mean time to failure ($T_f$) to the sum $T_f+T_c/2+T_r$ of the mean time to failure and the mean time to repair ($T_c/2+T_r$). However, in the context of checkpointing, the useful time in the mission application is $T_f - \frac{T_f T_s}{T_c}$. Therefore, availability in this context is given by Eqn. \ref{eq:AV}
\begin{equation}
Availability(T_f,T_s,T_c,T_r) = \frac{T_f-\frac{T_f T_s}{T_c}}{T_f+T_c/2+T_r} \label{eq:AV} \\
\end{equation}
As was stated before, in data center operations availability is the most important metric therefore optimal checkpoint interval must be chosen to maximize it. Differentiating Eqn. (\ref{eq:optAv}) we get
\begin{equation}
\frac{\partial Availability(T_f, T_s, T_c, T_r)}{\partial T_c} = \frac{\frac{T_f T_s}{{T_c}^2}}{\frac{T_c}{2}+T_f+T_r}+\frac{\frac{T_f T_s}{T_c} - T_f}{2(\frac{T_c}{2}+T_f+T_r)^2}==0 \label{eq:pAv} \\
\end{equation}
Solving Eqn. (\ref{eq:pAv}) we the optimal checkpoint interval in Eqn. (\ref{eq:optAv})
\begin{equation}
T_c = T_s + \sqrt{2(T_f+T_r)T_s+{T_s}^2} \label{eq:optAv} \\
\end{equation}

Equation (\ref{eq:optAv}) is different from Eqn. (\ref{eq:optLT}); however, for $T_s << T_r \And T_r << T_f$ asymptotically converges to $\sqrt{2 T_f T_s}$. For example, with $T_f = 1$ hour, $T_s=1$ second and $T_r = 4$ minutes; lost time optimal checkpoint interval, expressed in minutes, $T_c$ = $\sqrt{2\times 1\times 60 \times 1 /60} = \sqrt{2} = 1.414...$
Likewise, the calculation of availability optimal checkpoint interval, using Eqn. (\ref{eq:optAv}), will give $T_c = 1.477...$ minutes.
Table \ref{tab:1} lists the lost time optimal and availability optimal checkpoint interval values for various combinations of $T_f, T_s \And T_r$. Table \ref{tab:1} shows that with weaker satisfaction of the inequalities, $T_s << T_r \And T_r << T_f$, results in more significant difference between lost time optimal and availability optimal checkpoint intervals.

\begin{table}[]
\caption{Lost Time Optimal $T_c$ (mins) \& Availability Optimal $T_c$ (mins)} 
\label{tab:1}
\centering
\begin{tabular}{|r|r|r|r|r|}
\hline
\multicolumn{1}{|c|}{$T_f$(hours)} & \multicolumn{1}{c|}{$T_s$ (secs)} & \multicolumn{1}{c|}{$T_r$(mins)} & \multicolumn{1}{c|}{Lost Time Optimal} & \multicolumn{1}{c|}{Availability Optimal} \\ \hline
1                                & 1                              & 4                              & 1.41                       & 1.48                       \\ \hline
1                                & 1                              & 16                             & 1.41                       & 1.61                       \\ \hline
1                                & 30                             & 4                              & 7.74                       & 8.52                       \\ \hline
1                                & 30                             & 16                             & 7.74                       & 9.23                       \\ \hline
2                                & 1                              & 4                              & 2.00                       & 2.04                       \\ \hline
2                                & 1                              & 16                             & 2.00                       & 2.14                       \\ \hline
2                                & 30                             & 4                              & 10.95                      & 11.66                      \\ \hline
2                                & 30                             & 16                             & 10.95                      & 12.17                      \\ \hline
\end{tabular}
\end{table}
   
\section{Error Detection Latency $T_e$}
Implicit in the analysis and derivation of optimal checkpoint interval is the assumption that the error detection latency $T_e$ is significantly less than $T_c$. This assumption is mostly true when the compute system has built-in \emph{hardware} (HW) error detection mechanisms. However, there are scenarios where the notification of messages (like \emph{i-am-alive} heart beat events in a distributed compute system) happen at an interval greater than $T_c$. The application may also have some extra periodic checks to increase the fault detection coverage and to minimize application performance overhead the frequency of these checks could be once in single digit hours. So in a situation where $T_e > T_c$, the correct formulation of $LostTime()$ and $Availability()$ is in the equations listed below:
\begin{equation}
 LostTime(T_f, T_s, T_c, T_r, T_e) = \lfloor {\frac {T_f}{T_c}} \rfloor T_s + \lfloor{\frac{T_e}{T_c}}\rfloor T_c + \frac{T_c} {2}+ T_r \label{eq:LT1} \\
\end{equation}

\begin{equation}
Availability(T_f,T_s,T_c,T_r, T_e) = \frac{T_f-\lfloor{\frac{T_f}{T_c}}\rfloor{T_s}}{T_f+\lfloor{\frac{T_e}{T_c}}\rfloor T_c + T_c/2+T_r}\label{eq:AV1} \\
\end{equation}

\begin{figure} 
     \centering
     \includegraphics[width=\textwidth]{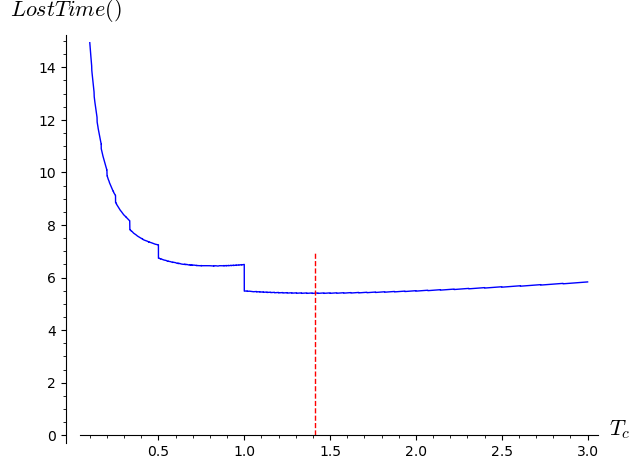}
     \caption{$LostTime()$ Optimality at $T_c = \sqrt{2}\ Minutes$: $T_f = 1\ hour, \\ T_s=1\ second, T_e =1\ minute \And T_r=4\ minutes$}
     \label{pict1}
\end{figure}    
\begin{figure} 
     \centering
     \includegraphics[width=\textwidth]{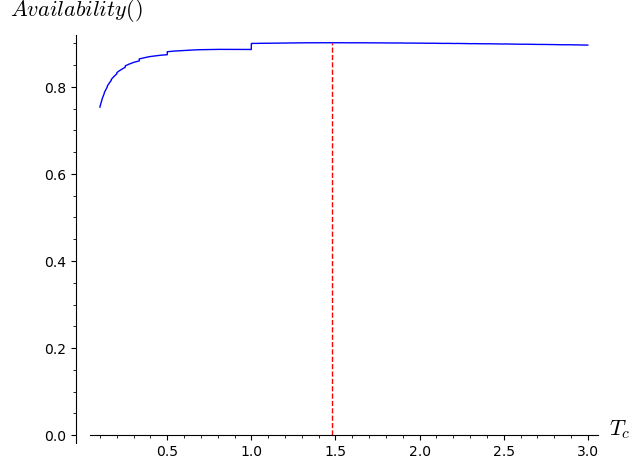}
     \caption{$Availabilty()$ Optimality at $T_c = 1.477...\ Minutes$: $T_f = 1\ hour, \\ T_s=1\ second, T_e =1\ minute \And T_r=4\ minutes$}
     \label{pict2}
\end{figure}    
\begin{figure} 
     \centering
     \includegraphics[width=\textwidth]{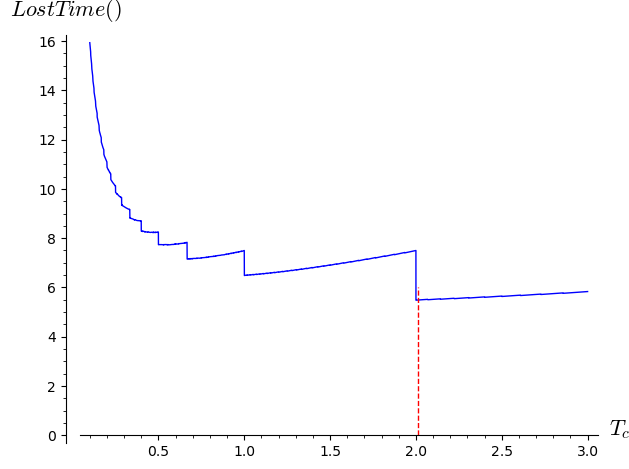}
     \caption{Shift in $LostTime()$ Optimality when $T_e > T_c$: $T_f = 1\ hour, \\ T_s=1\ second, T_e =2\ minutes \And T_r=4\ minutes$}
     \label{pict3}
\end{figure}
\begin{figure} 
     \centering
     \includegraphics[width=\textwidth]{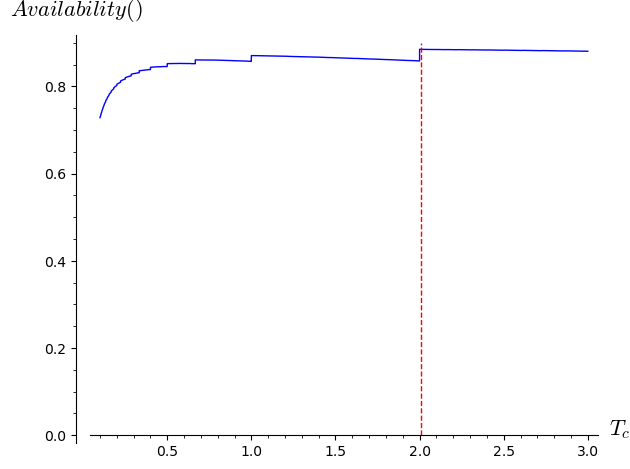}
     \caption{Shift in $Availability()$ Optimality when $T_e > T_c$: $T_f = 1\ hour, \\ T_s=1\ second, T_e =2\ minutes \And T_r=4\ minutes$}
     \label{pict4}
\end{figure}    
 
 Equations (\ref{eq:LT1}) and (\ref{eq:AV1}) have discontinuities due to the $floor$ terms and therefore will not have well-defined partial derivatives with respect to the $T_c$ variable.  This means conventional method of solving the partial derivative expression equal to 0 will not work. Please note that even with $T_e < T_c$, equations (\ref{eq:LT1}) and (\ref{eq:AV1}) will be different from the corresponding equations (\ref{eq:LT}) and (\ref{eq:AV}) due to the $\lfloor{T_f/T_c}\rfloor$ term. Plots represented by Figures \ref{pict1} and \ref{pict2}, using equations  (\ref{eq:LT1}) and (\ref{eq:AV1}), show that the optimal checkpoint intervals derived using equations (\ref{eq:LT}) and (\ref{eq:AV}) hold true.

However, when we consider $T_e > T_c$, the checkpoint intervals, lost time optimal and availability optimal, show a significant departure from the ones derived from equations (\ref{eq:LT}) and (\ref{eq:AV}). Figures \ref{pict3} and \ref{pict4} illustrate this. It turns out the optimal check point interval is slightly greater than $T_e$. For example, plots in Figures \ref{pict3} and \ref{pict4} show optimal checkpoint interval slightly above 2 minutes.       

 \section{Conclusions}
 
One key takeaway from the error detection latency discussion is the need to save more than one snapshot of the checkpoint state. Even when $T_e$ is significantly smaller than $T_c$, there is always a corner case when the error is detected during the checkpoint state save event. This corner case invalidates the most recently saved checkpoint state and requires rollback recovery to pick a checkpoint state before the most recent one. The case where $T_e$ is greater than any established optimal checkpoint interval would require a new checkpoint interval for the class of errors with longer detection latency. In a data center comprising tens of thousands of nodes, checkpoint state save at periodic intervals puts enormous burden on the network bandwidth and storage bandwidth and capacity. Therefore, we need a careful management of the number of checkpoints and the optimality of checkpoint interval as a function of error latency categories.

In summary, we formulated a simplified derivation to the lost time optimal checkpoint interval. This is identical to that in \cite{young_1974}. In this paper, for the first time, we made the observation that the most important performance metric in the data center is system availability. While minimizing lost time is an important contributor in maximizing availability, from the standpoint of optimal checkpoint interval the objective function that is relevant is availability. We show that availability-optimal $T_c = T_s + \sqrt{2(T_f+T_r)T_s+{T_s}^2}$. For $T_s << T_r \And T_r << T_f$, availability-optimal checkpoint interval is asymptotically same as lost-time-optimal checkpoint interval.
    
\bibliographystyle{ieeetr}

\bibliography{checkpoint}

\end{document}